# Gas Pixel Detectors for X-ray Polarimetry applications

R. Bellazzini[a], F. Angelini[b], L. Baldini[a], F. Bitti[a], A. Brez[a], F. Cavalca[a], M. Del Prete[a], M. Kuss[a], L. Latronico[a], N. Omodei[a], M. Pinchera[a], M. M. Massai[b], M. Minuti[a], M. Razzano[a], C. Sgro'[a], G. Spandre[a], A. Tenze[a], E. Costa[c], P. Soffitta[c]

[a]INFN sez.Pisa, Largo B. Pontecorvo, 3 I-56127 Pisa, Italy
[b]University of Pisa and INFN-Pisa, Largo B. Pontecorvo, 3 I-56127 Pisa, Italy
[c]Istituto di Astrofisica Spaziale e Fisica Cosmica of INAF, Via del Fosso del Cavaliere, 100, I-00133 Roma, Italy

**Abstract**

We discuss a new class of Micro Pattern Gas Detectors, the Gas Pixel Detector (GPD), in which a complete integration between the gas amplification structure and the read-out electronics has been reached. An Application-Specific Integrated Circuit (ASIC) built in deep sub-micron technology has been developed to realize a monolithic device that is, at the same time, the pixelized charge collecting electrode and the amplifying, shaping and charge measuring front-end electronics. The CMOS chip has the top metal layer patterned in a matrix of 80 μm pitch hexagonal pixels, each of them directly connected to the underneath electronics chain which has been realized in the remaining five layers of the 0.35μm VLSI technology. Results from tests of a first prototype of such detector with 2k pixels and a full scale version with 22k pixels are presented. The application of this device for Astronomical X-Ray Polarimetry is discussed. The experimental detector response to polarized and unpolarized X-ray radiation is shown. Results from a full MonteCarlo simulation for two astronomical sources, the Crab Nebula and the Hercules X1, are also reported.

1. Introduction

Many astronomical objects, such as active galactic nuclei or spinning bodies (pulsars and black holes) emit polarized radiation at X-ray wavelengths. Astronomers believe that measuring this polarization will provide fundamental information about the geometry and internal structure of these sources allowing to uncover how matter behaves in extremely intense magnetic and gravitational fields. Conventional Polarimeters based on Bragg diffraction or Thompson scattering methods are characterized by a poor sensitivity and have given to date positive results only for a very few bright sources, as the first and only generally-accepted measurement of the Crab nebula. This measurement has been made with a Bragg crystal polarimeter more than 30 years ago[1,2]. To capture the polarization of faint and weakly polarized sources we have developed a new instrument based on the photoelectric effect, a process very sensitive to photon polarization and with a large cross-section in the low energy range (2÷10 keV) of great astronomical interest. This instrument belongs to the class of Micro Pattern Gas Detectors of which it represents the latest stage of development. To derive the polarization of the X-ray photon the few hundreds micron track of the photoelectron, which is emitted mainly in the direction of the photon electric field (polarization vector) is reconstructed by a finely structured collecting electrode. In our case this is the top metal layer of a VLSI chip realized in 0.35 μm CMOS technology. This layer has been patterned as a honeycomb pixel array with a pitch of 80 μm. Each single pad acts as individual charge collecting electrode and it is connected to a full electronics chain (pre-amplifier, shaping amplifier, sample and hold, multiplexer) built immediately below it, in the remaining five layers of the CMOS technology. The high granularity of the pixel array allows to achieve at the same time true 2D imaging capability and high rate operation. Two versions of the VLSI chip have been designed and built: a first one with 2101 pixels[3] and a successive upgrade to 22000 pixels[4]. Both have been assembled with a gas charge amplifying electrode (a fine pitch GEM foil) and the enclosure drift window (25μm Mylar foil) to form the Gas Pixel Detector. The GEM has a standard thickness of 50μm with holes, on a triangular pattern, of 50μm diameter at 90μm pitch. The drift region (*absorption gap*) is 6 mm while a thin spacer defines a 1mm thick *collection gap* between the bottom GEM and the pixel matrix of the read-out chip. A description of the instrument, of the VLSI read-out chip and the results of laboratory tests obtained with a unpolarized 5.9 keV X-ray source and a ~100% polarized 5.4 keV source are presented in the next sections. In the last section results of the Monte Carlo simulation of the detector response as polarimeter to few celestial sources are shown.

## 2. The Gas Pixel Detector: principle of operation

Fig.1 explains schematically the concept of the Gas Pixel Detector for X-ray detection. The photon converts in a low Z gas mixture (usually Neon 50%-DME 50%) emitting a photoelectron that produces an ionization track in the gas. The electrons of the track are drifted toward an amplification electrode, the Gas Electron Multiplier (GEM), where they are multiplied and then collected by the underneath pixelized read-out plane.

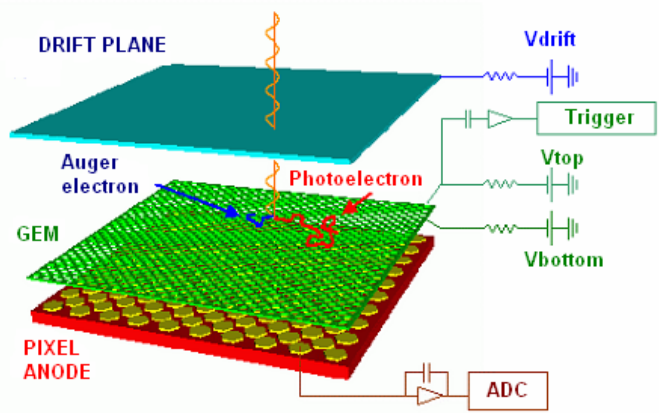

Fig.1. The concept of the Gas Pixel Detector.

The introduction of the GEM as a charge amplification structure separated from the read-out plane of a gas detector, has opened the possibility to freely pattern the collecting electrode in a fine multi-pixel array. In the first prototype we developed[5] this was obtained routing out the signal from each pad to an external electronic channel. Technological constraints limit severely the dimension of the fan-out (number and length of connection lines) to the front-end electronics. Going to pixel size smaller than 100 μm and number of channels greater than 1000 becomes unfeasible. Furthermore, cross-talk between channels and noise level due to high input-capacitance is not negligible. To overcome all these limitations we have designed and fabricated a custom CMOS ASIC. The first generation we have produced, (ASIC-I), had 2101 read-out pixels at 80 μm pitch, each one individually read by a full chain of analog electronics. Due to the very low pixel capacitance, a noise value of 100 electrons rms has been measured for such device. With typical electric fields of 1kV/cm in the drift region and 80÷100kV/cm through the GEM, the detector operates at gas gain of few thousands. In these conditions with a threshold at 3 sigma noise (300 electrons) a non negligible single primary electron sensitivity is reached. An example of a real track produced in the gas by the photo-conversion of a 5.9 keV X-ray and recorded by the detector, is shown in fig.2.

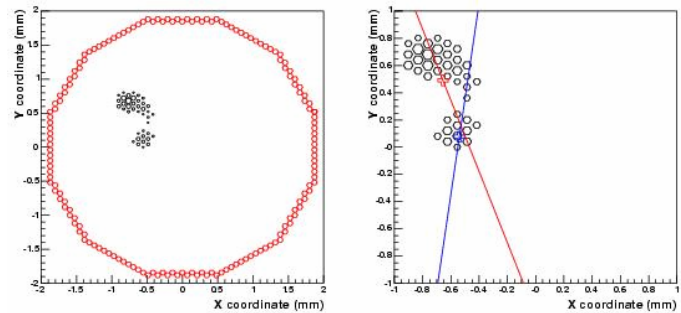

Fig.2. Real track produced in the gas by a 5.9 keV photon. The reconstruction algorithm develops in the following steps: 1) barycenter evaluation of the charge distribution (red cross), 2) reconstruction of the principal axis direction (red line), 3) conversion point evaluation (blue cross), 4) emission direction reconstruction (blue line). The polarization is derived from the photoelectrons angular distribution.

### 2.1. ASIC-II: the 22k VLSI chip

Still realized in the same 0.35μm CMOS technology, ASIC-II has an active area increased of roughly a factor ten with respect to ASIC-I; that is from ~12mm$^2$ (4mm diameter) of the first prototype to the actual 11×11mm$^2$, corresponding to a total number of 22080 pixels (fig.3).

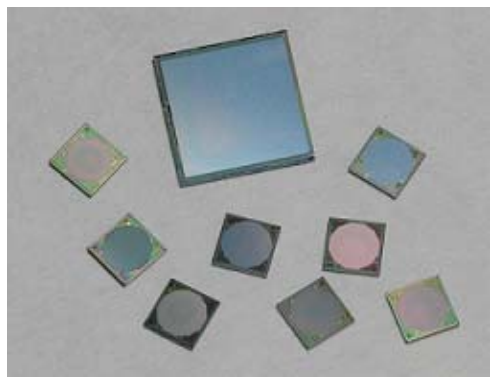

Fig.3. Photo of the bare ASIC prototypes showing the large increase in active area of the second generation chip (22080 pixels) respect to the first one (2101 pixels).

The pixel elements are still hexagonally shaped but arranged according to a honeycomb pattern on a squared area (fig.4). Each pixel has the same conditioning chain for the signal (charge sensitive amplifier and shaping circuit) as in the first prototype. The circuit is organized in 8 identical clusters of 2760 pixels (20 rows, 138 pixels each) each one with an independent differential analog readout buffer. Upon the activation of an external digital control input (*MaxHold*) set by the trigger signal from the top GEM electrodes, the peak detection of the shaped pulse is initiated and the maximum is stored inside each pixel cell for subsequent readout. The *hold* mode is terminated when a pulse is applied to the (*AnaReset*) input. In this way all pixels of the matrix return to the track mode simultaneously.

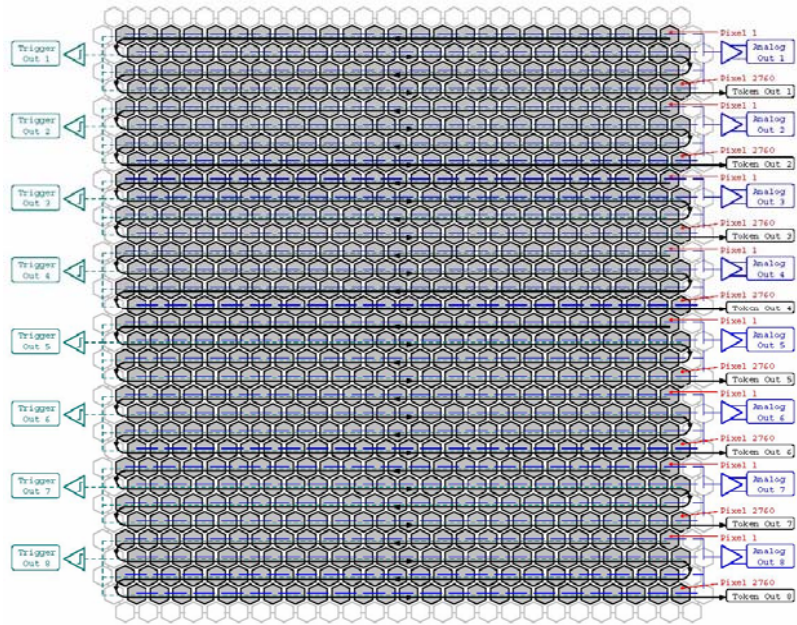

Fig.4. ASIC simplified pixel layout and serial readout architecture (actual number of pixels is larger than shown).

**Timing Characteristics**
VDD-VSS ≥ 3.0V unless otherwise noted.

| Symbol | Parameter | Min | Typ | Max | Unit |
|---|---|---|---|---|---|
| Tpk | Shaper peaking time | 3 | 4 | 5 | us |
| Tdh | Delay between incoming event & rising edge of MaxHold | 0 | | 2 | us |
| Tmh | MaxHold pulse width | 1 | | Tstore | us |
| Tpd | Peak detection mode duration | 10 | | 20 | us |
| Tstore | Analog data retention time for a drift during hold mode lower than 0.1fC equivalent input charge) | 1 | | | ms |
| Tar | AnaReset pulse width | 1 | | | us |
| Tr | Analog memory recovery time after hold | | | 100 | us |

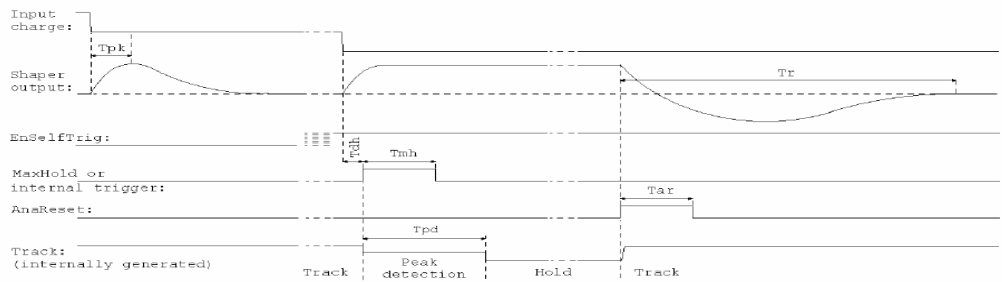

Fig.5. Readout timing characteristics and definition).

Fig.5 shows the timing characteristics of the chip. The readout is accomplished by sequentially connecting the output of each cell to the analog bus common to each cluster. For this purpose each pixel includes a shift register element, which can also be used for electrical test and calibration. The chip has been tested at 5MHz read-out frequency corresponding to a frame rate (read-out time per cluster) of 550μs, but it can nominally work up to 10 MHz. In these conditions a source trigger rate of 1÷2kHz is sustainable. Clock drivers, bias circuitry, trigger output & analog buffer are placed on the left- and right-hand side of the chip. Fig.6 shows a photo of both chips bonded to their packages (left) and, on the right, an enlarged view on the 22k ASIC-II. For the chip control and read-out a very compact DAQ system has been designed. It is implemented on three boards stacked together. The detector and the trigger electronics chain are mounted on the outermost board (*SENSOR_BOARD*).

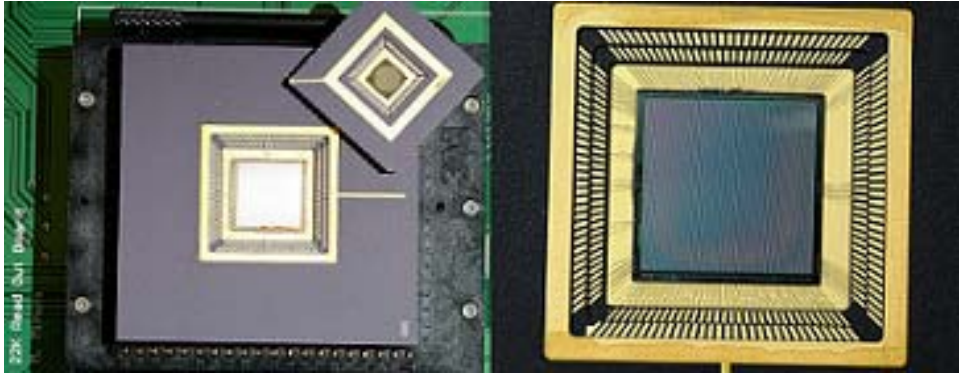

Fig.6. Photo of the two prototypes bonded to their ceramic package (left). Close-up of the 22k pixels chip (right).

The amplifying chain for the trigger signal from the top GEM electrode is based on the AMPTEK A206 Charge Sensitive and Shaping Amplifier followed by a threshold discriminator (Maxim 903CSA). The GEM signal reaches the peak in few hundreds nanoseconds, a time which is significantly shorter than the peaking time of the analog signal from the pixels (3÷4 μs).

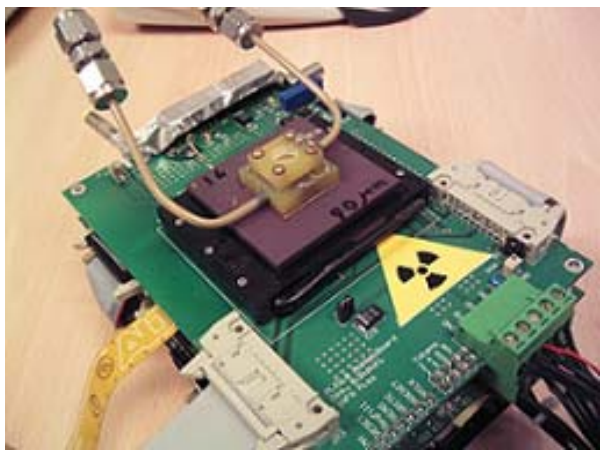

Fig.7. Photo of the large area ASIC-II mounted on the control motherboard and covered by the top section of the detector. The gas-tight enclosure glued on top of the chip is formed, from top to bottom, by: (1) entrance window, (2) absorption gap spacer, (3) GEM foil, (4) collection gap spacer.

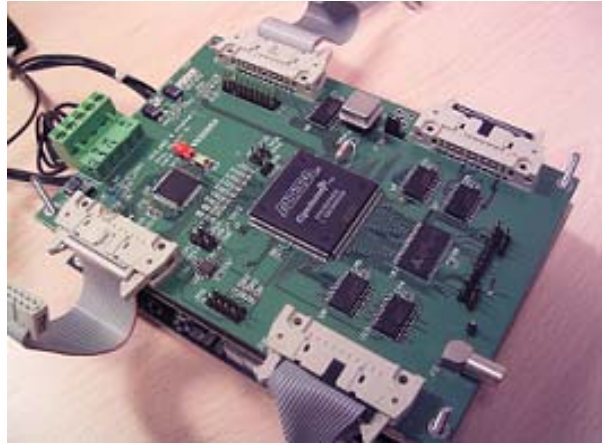

Fig 8. The *SEQ_ADC_BOARD*. The Altera FPGA used to generate control signals for the ASIC-II chip is recognizable at center of the board.

The Interface Board at the stack bottom implements on Altera FPGA an embedded RISC processor NIOS II for high level operations. An Ethernet controller on the board is used to establish 100Mbps TCP/IP connection to the PC for data transfer. A graphic interface for instrument control and data acquisition has been developed in LabVIEW. The VI performs a bidirectional communication with the DAQ device through a TCP network connection at the specified IP address and port.

## 2.2. Laboratory tests

A complete set of tests has been performed on the chip. A noise level of less than 2 ADC counts (rms) corresponding to an ENC of ~100 electrons has been measured (fig.9) as in the first prototype.

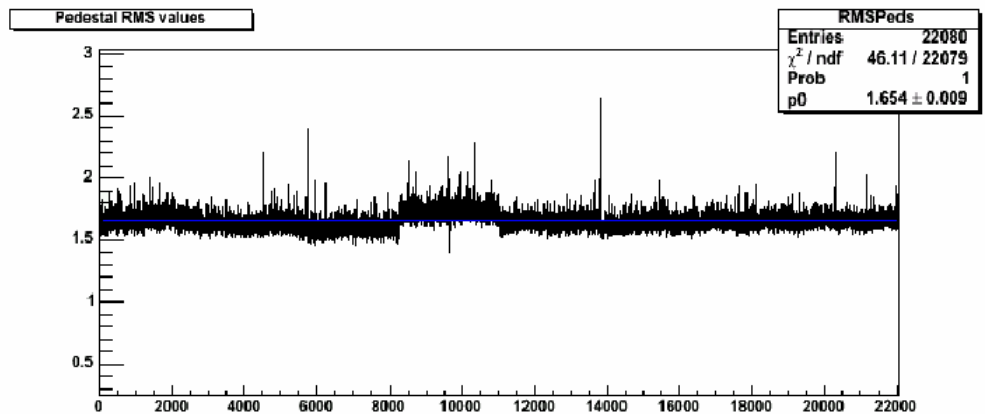

Fig.9. Average noise for all the 22k pixels. All pixels are working. Amplifier input sensitivity = 100 ADC counts/fC.

Fig.10 shows the raw data (pedestal subtracted) and the corresponding reconstructed tracks in the case of an event with multiple tracks. Three 5.9 keV photons from a strong $^{55}$Fe source converted in the gas within the peak search time window (~10 μs). An enlarged view of the track with the highest total charge is also shown. Information about the degree and angle of polarization of the incident radiation is derived from the angular distribution of the initial direction of the photo-electron. This distribution is modulated as $\cos^2(\varphi)$, where $\varphi$ is the linear polarization angle.

The lower limit on the Minimum Detectable Polarization (MDP) is set by the residual modulation due to systematic effects (if any), measured when observing a totally unpolarized X-ray source. By using a $^{55}$Fe X-ray source we have obtained (fig.11) a best fit residual modulation of 0.59% with a statistical error of ± 0.81%, from a run containing ~30k events. This result is fully compatible, within the statistical error limit, with the absence of any modulation. Obviously, setting a more stringent limit on the systematic error will require a much larger data sample.

A measure of the modulation factor for polarized photons has been carried out by using radiation from a Cr X-ray tube (20 kV, 35 mA). The X-ray beam is Thompson scattered through a Li target (6mm in diameter, 70mm long), canned in a beryllium case (500 µm thick) in order to prevent oxidation and nitridation from air[6]. The geometry of the output window of the scatterer and the distance with respect to the detector limit the scattering angles to ~90º so that the radiation impinging the detector is highly linearly polarized. A typical modulation factor of ~48% has been measured when the source is polarized better than 98%. The sensitivity of the system to changes in polarization angle has been tested simply by rotating the detector around the vertical direction of a small angle (few degrees) clockwise (a) and counter-clockwise (b) respect to the initial position (90º, see fig.12). A corresponding change in the reconstructed polarization angle has been verified and angles of 70º±1º, and 102º±1º have been measured, respectively (fig.13).

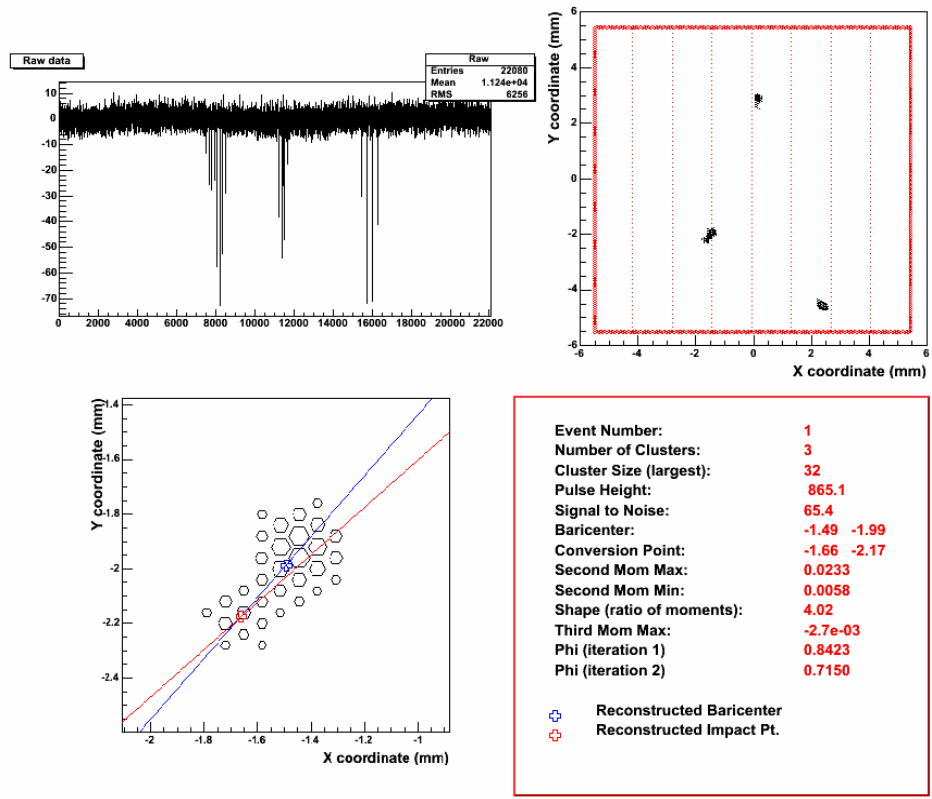

Fig.10. Raw data (top-left panel) and reconstructed tracks (top-right panel) from 5.9 keV photons. The track with the highest total charge is shown separately (bottom-left panel). The relevant parameters derived from the analysis are also reported (bottom-right panel).

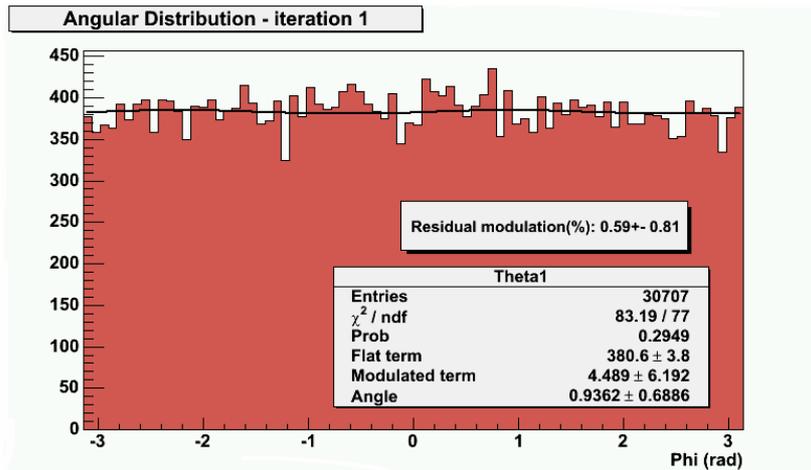

Fig.11. Polarimeter response to an unpolarized $^{55}$Fe X-ray source. The residual modulation is 0.59% ± 0.81%. The plot contains also the fit parameters.

In fig. 14 some "real" tracks obtained by irradiating the detector with photons from the X-ray Cr tube, are shown. The very good imaging capability of the detector has been tested by illuminating from above with a $^{55}$Fe source a small pendant (few mm in size) placed in front of the detector. The 'radiographic' images obtained by plotting the barycenters and the conversion points are reported in fig.15.

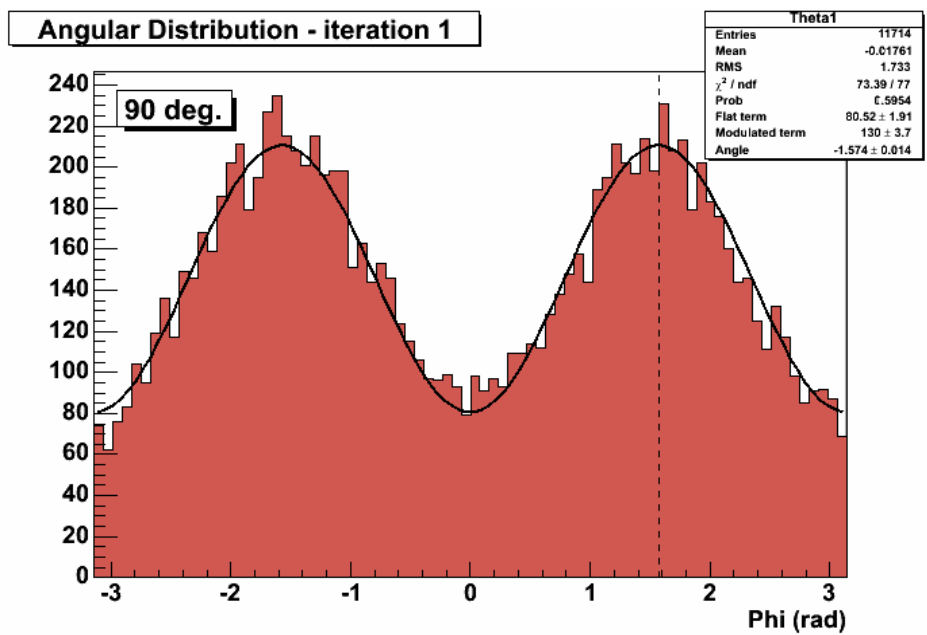

Fig.12. Angular distribution of the reconstructed photoelectron tracks. The polarization angle is 90°±1°. For this specific measurement the modulation factor obtained in a slightly different source configuration is ~45%.

It is worth to note the large improvement in image reconstruction when using the absorption point instead of the barycenter in case of "light" gas mixture (p.e. Ne 80% - DME 20%) with respect to high Z gas mixtures as, for example, standard Ar-based mixtures (fig.16). In this latter case the barycenter is very close to the conversion point, while for Ne-based mixture it can be quite far from it.

It should be also underlined that after several months of intensive operation, no single pixel has been lost for electrostatic or GEM discharges, or for any other reason.

To further push forward the track reconstruction capability at very low photon energies, 1÷2 keV (hence very small track lengths), where the photon flux from astronomical sources is higher and the mirror effective area is wider, a third ASIC generation has been designed and built with~ 100000 pixels at 50 μm pitch.

The VLSI chip (in 0.18 μm CMOS technology) integrates more than 16.5 million transistors and it is organized as a 15mm×15mm active area. It is subdivided in 16 identical clusters, each one with an independent differential analog readout buffer. Each cluster has a customizable internal self-triggering capability with independently adjustable thresholds. An on-chip wired-OR combination of each cluster self-triggering circuit holds the maximum of the shaped signal on each pixel. The self-triggering function also includes an on-chip signal processing for automatic localization of the event coordinates. In this way it will be possible a significant reduction of the readout time by limiting the signal output only to the pixels belonging to the region of interest.

The chip is now under test in our laboratory and results will be presented soon.

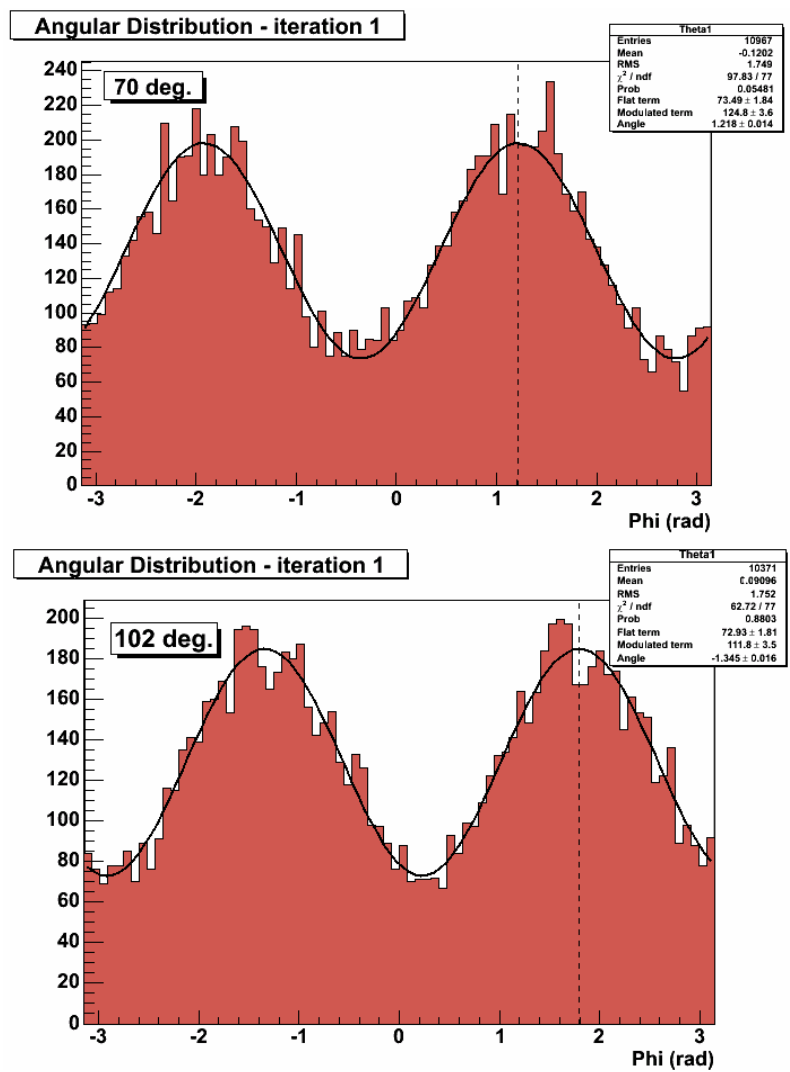

Fig.13. Changes of the reconstructed polarization angle after a rotation of the detector of 20º clockwise (a) and 12º counter-clockwise (b) respect to the original position.

## 3. MonteCarlo simulation and results

All the physics processes ruling the operation of this detector as X-ray polarimeter have been completely Monte Carlo simulated. These processes include the photoelectric interaction, the scattering and slowing of the primary electrons in the gas, drift and diffusion, gas multiplication and the final charge collection on the readout plane.

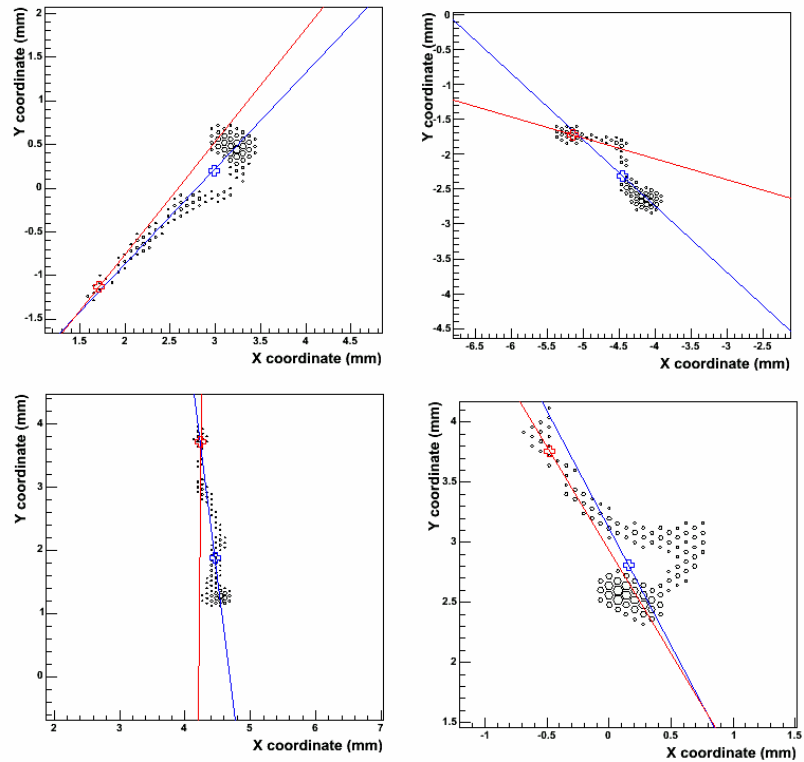

Fig.14. Real tracks obtained by irradiating the detector with photons from the X-ray Cr tube (gas mixture Ne 80% - DME 20%) Red and blue lines are explained in the caption of fig.10.

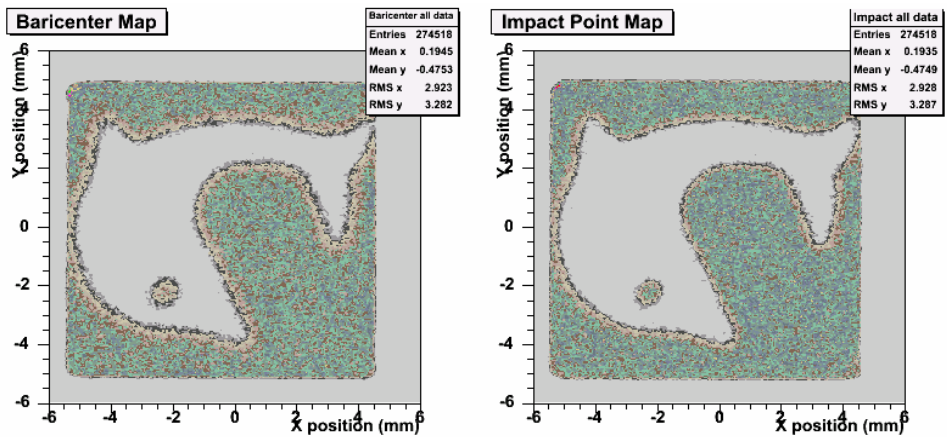

Fig.15. *Radiographic* image of a small pendant obtained with 5.9 keV photons from a 55Fe source.

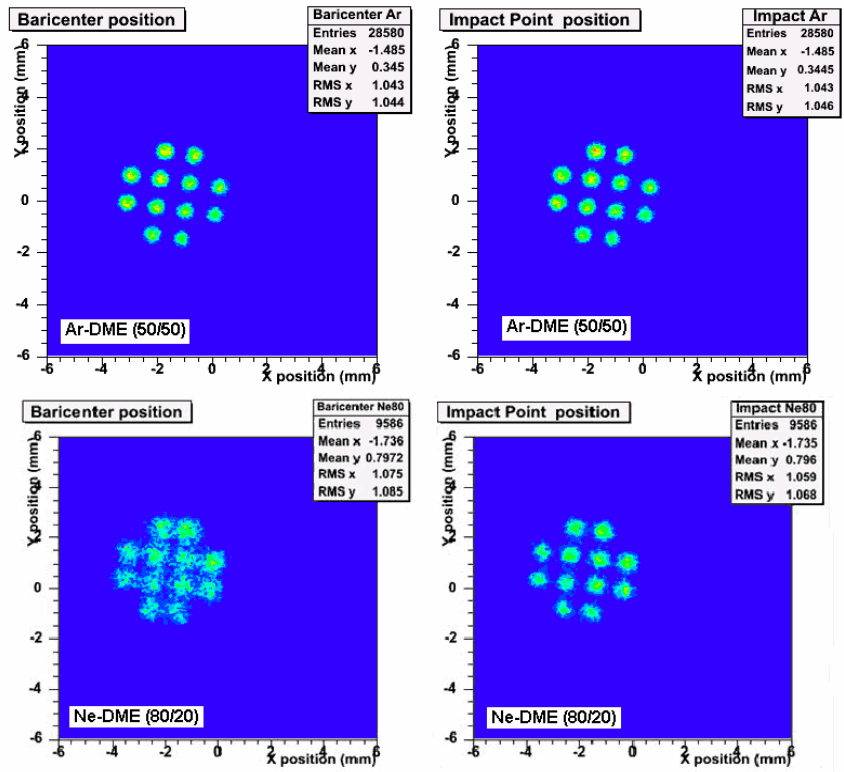

Fig. 16. Differences in image reconstruction using barycenters or conversion points (impact Point in figure) in a Ar-DME mixture (top plots) and in a lighter Ne-DME mixture (bottom plots). Holes have 0.6 mm diameter, 2 mm apart.

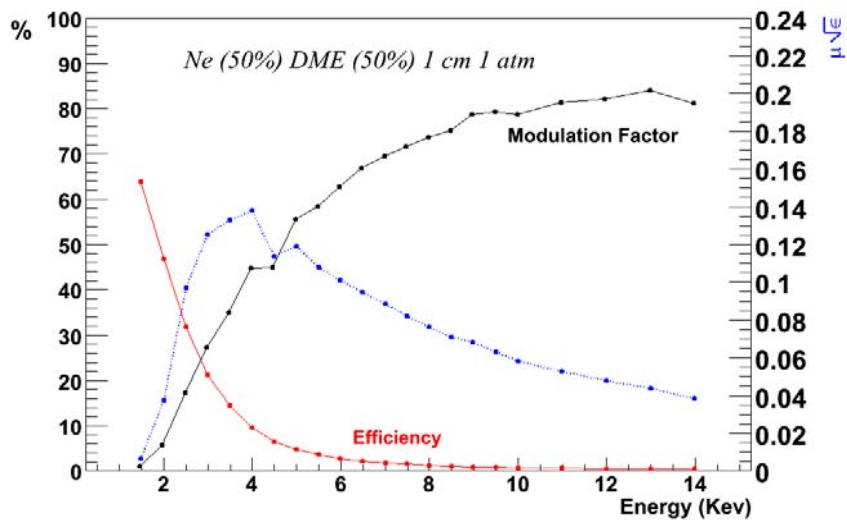

Fig. 17. Efficiency and modulation factor vs. photon energy in Ne 50%-DME 50%.

All of them are function of the photon energy and of gas parameters such as composition, pressure and drift path. Description of the computational model can be found elsewhere[7,8].

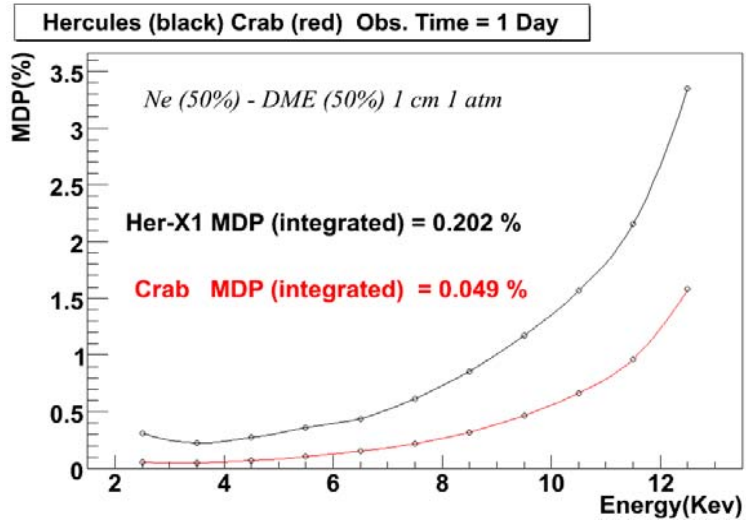

Fig.18. Minimum Detectable Polarization as a function of the photon energy for the proposed polarimeter at the focus of XEUS optics for two bright X-ray sources.

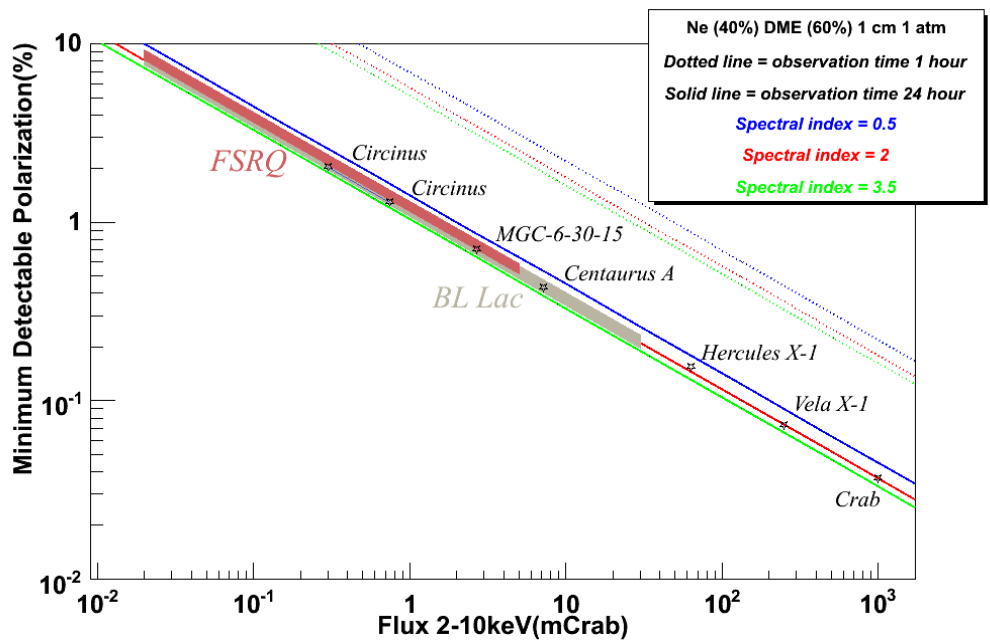

Fig.19. Minimum Detectable Polarization for the proposed Polarimeter as a function of the flux, for a few representative sources.

An extensive study of fundamental parameters such as the modulation factor and the detection efficiency that, together with the mirror effective area, the observation time and the celestial source flux, determine the sensitivity of the polarimeter, has been carried out with the simulator in a wide energy range (from 1 keV to few tens of keV) with different gas mixtures. As an example, fig.17 shows the modulation factor and the efficiency obtained with 1cm-1atm pressure of Neon 50% - DME 50%. The plot of the Polarimeter Quality Factor, defined as QF=$\mu\sqrt{\varepsilon}$ [7], is also shown (dotted line). The resulting MDP is plotted in fig.18, assuming the actual polarimeter at the focus of the optics of XEUS (X-Ray Evolving Universe Spectrometer), an ESA permanent spaceborne X-ray observatory planned to be launched around 2015.

Fig.19 summarizes the sensitivity obtained for different exposure times and for a few representative sources. With observations of one day we can measure the polarization of several AGNs down to few % level. For its high sensitivity this detector has been proposed at the focus of a large area telescope as those ones foreseen for the New Generation X-ray Telescope in the frame of the ESA Cosmic Vision 2015-2025.

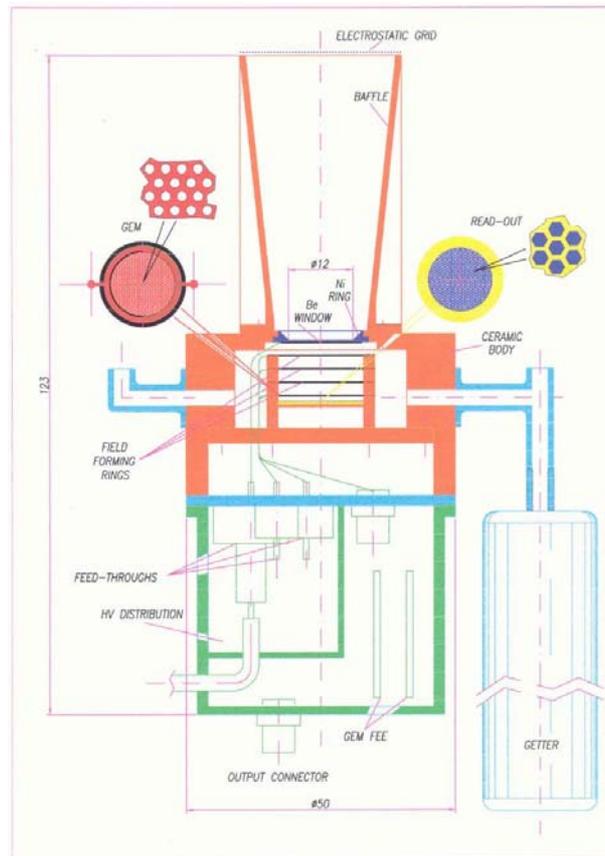

Fig.20. XPOL: a possible design based on established technology.

A possible design of this instrument (named XPOL) based on established technology is shown in fig.20. Critical parts as GEM and VLSI have already reached a high degree of development, while detector body, Be window, gas handling, HV, etc. are well established technology. Including HV Power Supply and control electronics the needed resources will be of the order of 10 kg and ~15 watts. The device is very compact and no cryogenics, nor rotations will be necessary.

## 4. Conclusions

The performance of the tested prototypes looks like a significant step forward, compared with traditional X-ray polarimeters and promises a large increase in sensitivity. In its final configuration the target performance of the device is the detection of ~1% polarization for few milli-Crabs sources (in the XEUS focal plane, for example). This sensitivity will likely allow polarimetry measurements to be made on thousands of galactic and extragalactic sources: a real breakthrough in X-ray astronomy. The final design with 100 k channels and 50 μm pixel size will bring the Gas Pixel Detector to the same level of integration of solid state detectors. Moreover, depending on pixel and die size, electronics shaping time, analog vs. digital read-out, counting vs. integrating mode, many other applications, than X-ray Polarimetry can be envisaged with this device. In this respect, it is worth noticing that following a similar approach, a digital counting chip developed for medical applications (Medipix2) has been shown to work when coupled to GEM or Micromegas gas amplifiers for TPC application at the next generation of particle accelerators ([9]).